\documentclass[12pt,a4paper]{article}
\usepackage{amsmath,amssymb,mathrsfs,framed,esint,slashed}
\usepackage[colorlinks]{hyperref}
\usepackage{color}
\usepackage[all]{xy}

\newcommand{\rem}[1]{}

\newcommand{\de}{{\rm d}}

\newcommand{\bq}{{\mathbf{q}}}

\newcommand{\bp}{{\mathbf{p}}}

\newcommand{\bx}{{\mathbf{x}}}

\newcommand{\bz}{{\mathbf{z}}}

\newcommand{\bE}{{\mathbf{E}}}
\newcommand{\bB}{{\mathbf{B}}}

\newcommand{\br}{{\boldsymbol{r}}}

\newcommand{\beq}{\begin{equation}}
\newcommand{\eeq}{\end{equation}}

\begin{document}

\title{Momentum maps for mixed states\\in quantum and classical mechanics}
\author{Cesare Tronci
\\
\vspace{-.1cm}
\footnotesize
\it Department of Mathematics, University of Surrey, Guildford GU2 7XH, United Kingdom\\
\footnotesize
\it Mathematical Sciences Research Institute,  Berkeley, CA 94720, United States\vspace{.1cm}}
\date{\small\sf For Darryl Holm, on the occasion of his 70th birthday}
\maketitle

\begin{abstract} 
This paper presents the momentum map structures which emerge in the dynamics of mixed states.  Both quantum and classical mechanics are shown to possess analogous momentum map pairs associated to left and right group actions. In the quantum setting, the right leg of the pair identifies the Berry curvature, while its left leg is shown to lead to  different realizations of the density operator, which are of interest in quantum molecular dynamics. Finally, the paper shows how alternative representations of both the density matrix and the classical density are equivariant momentum maps generating new Clebsch representations for both quantum and classical dynamics. Uhlmann's density matrix \cite{Uhlmann} and Koopman wavefunctions \cite{Koopman} are shown to be special cases of this construction.
\end{abstract}

\bigskip 

{\footnotesize

\tableofcontents

}

\section{Pure vs. mixed states: quantum and classical}

The geometric setting of quantum mechanics has been attracting attention ever since the work of Kibble \cite{Kibble}, who showed how Schr\"odinger's equation is a Hamiltonian system on the projective Hilbert space. Over the years, the geometric viewpoint of both pure and mixed states in quantum mechanics has been developed in several works \cite{Anandan, Brody, CaIbMaMo, deGosson, Littlejohn2, Montgomery91, Uhlmann}. However, while the difference between pure and mixed quantum states is widely known, its classical correspondent is only rarely reported in the literature, see e.g. \cite{ChernoffMarsden,Shirokov}. This difference is especially important when one considers the coexistence of quantum and classical systems. For example, in quantum molecular dynamics, the complexity of a full quantum treatment requires approximations which treat parts of the molecule (the nuclei) as classical particles interacting with a pure state wavefunction governing the electronic quantum dynamics \cite{Baer, Marx}. 

This paper presents a correspondence between the geometric features underlying the dynamics of quantum and classical states in terms of momentum map structures. This is a relatively new perspective. Indeed, while the momentum map character of projection operators in quantum mechanics and point measures in the classical phase-space have long been known, a deeper investigation of the several other momentum maps appearing in quantum mechanics has begun only more recently. For example, in \cite{Sawicki1,Sawicki2} momentum maps were used for multipartite systems to characterize entanglement, while in \cite{BLTr14,BLTr15,OhTr} momentum maps were related to expectation value dynamics. On the other hand, in the case of multipartite systems, partial traces of the density matrix have  long been known to identify momentum maps \cite{Montgomery91}.

In this paper, the concept of momentum map is  applied to mixed states in both quantum and classical mechanics. While this section continues by reviewing the geometric setting of quantum and classical pure states, geometric extensions of the concept of quantum (and classical) mixture are presented later. In this generalized context, this paper shows that the celebrated Berry curvature \cite{berry1984quantal} also identifies a momentum map, whose dynamics appears in recent molecular dynamics models \cite{abedi2010exact,abedi2012correlated}  beyond the Born-Oppenheimer approximation \cite{Baer,born1927quantentheorie}. In the last part of the paper, new momentum map structures are shown to recover and extend alternative representations of both quantum and classical mechanics, such as Uhlmann's density matrix of quantum states \cite{Uhlmann} and the Koopman wavefunction for classical dynamics \cite{Koopman}.

\subsection{Quantum states}
Consider a physical system consisting of only one particle. In the quantum case, the particle dynamics may have two alternative descriptions depending on whether the system is in a pure or a mixed state. If the system is in a mixed state, then the particle dynamics is given in terms of a  positive-definite Hermitian operator $\rho$ defined on the quantum Hilbert space $\mathscr{H}$ and obeying the quantum Liouville equation
\beq\label{QLiouville}
i\hbar\partial_t\rho=[H,\rho]
\,,
\eeq
where $H$ is the Hermitian Hamiltonian operator for  the system. Notice that, since formally $\rho$ evolves under unitary transformations as $\rho(t)=e^{-itH/\hbar}\rho_0e^{itH/\hbar}$, both the unit-trace and the positivity conditions $\operatorname{Tr}\rho=1$ are simply added here as  initial conditions that are preserved in time. If we denote by ${\cal S}_Q\subset\operatorname{Her}(\mathscr{H})$ the convex subset of density operators in the set of Hermitian operators, its extreme points define the pure states. The latter are realized in terms of projection operators of the type
\beq
\rho=\psi\psi^\dagger
\,,
\label{PStates}
\eeq
where $\psi\in\mathscr{H}$ is the usual wavefunction satisfying the Schr\"odinger equation
\[
i\hbar\partial_t\psi=H\psi
\,.
\]
Typically, the wavefunction is normalized, so that $\|\psi\|^2=1$ and $\operatorname{Tr}\rho=1$. In this work, the normalization of the wavefunction is regarded as an initial condition that is preserved by the unitary dynamics produced by Schr\"odinger's equation. Another possibility would be to work directly on the unit sphere $S(\mathscr{H})$ in the Hilbert space $\mathscr{H}$ or the projective Hilbert space ${P}\mathscr{H}$ \cite{Kibble}. However, here we prefer to deal with the Hilbert space $\mathscr{H}$ itself to simplify the treatment.
In addition,  unless otherwise specified, in this paper we shall assume that the Hilbert space is finite-dimensional, that is $\mathscr{H}=\Bbb{C}^n$. Although most results formally apply also in the infinite-dimensional case (typically, $\mathscr{H}=L^2(\Bbb{R}^3$)), a finite-dimensional Hilbert space allows avoiding several important difficulties emerging in infinite dimensions; for example, see the discussions in \cite{deGosson2}.

It is well known \cite{CaIbMaMo} that the map
\[
\psi\mapsto -i\hbar\psi\psi^\dagger
\]
is an equivariant momentum map for the left representation $\psi\mapsto U\psi$ of operators  $U\in{\cal U}(\mathscr{H})$ in the unitary group ${\cal U}(\mathscr{H})$ on the quantum Hilbert space. 
This is easily seen by considering the canonical symplectic structure on $\mathscr{H}$, which is given as
\beq\label{sympform1}
\omega(\psi_1,\psi_2)=2\hbar\operatorname{Im}\langle\psi_1|\psi_2\rangle
\eeq
where $\langle\cdot|\cdot\rangle$ denotes the standard inner product, that is $\langle\psi_1|\psi_2\rangle=\operatorname{Tr}(\psi_1^\dagger\psi_2)$. The momentum map $J:\mathscr{H}\to\mathfrak{u}(\mathscr{H})^*$ is given by the usual formula \cite{MaRa2013,HoScSt2009} for linear Hamiltonian actions,  that is
\beq\label{momapformula1}
\langle J(\psi),\xi\rangle=\frac12\,\omega(\xi(\psi),\psi)=\hbar\operatorname{Im}\langle\xi\psi|\psi\rangle=\langle -i\hbar\psi\psi^\dagger,\xi\rangle
\,,
\eeq
where $\xi\in \mathfrak{u}(\mathscr{H})$ and $\xi(\psi)$ denotes the infinitesimal Lie algebra action of $\mathfrak{u}(\mathscr{H})$ on $\mathscr{H}$.
On the left-hand side of \eqref{momapformula1}, we have used the following notation for the real-valued pairing
\[
\langle \mu,\xi\rangle=\operatorname{Re}\big(\operatorname{Tr}(\mu^\dagger \xi)\big)
\,,
\qquad\qquad\forall\,\mu\in \mathfrak{u}(\mathscr{H})^*,\quad\ \forall\,\xi\in \mathfrak{u}(\mathscr{H})\,,
\]
and, as $\mathscr{H}$ is finite-dimensional, we have identified $\mathfrak{u}(\mathscr{H})^*\simeq\mathfrak{u}(\mathscr{H})$ via the inner product $\langle \mu|\xi\rangle=\operatorname{Tr}(\mu^\dagger \xi)$.

\subsection{Classical states}
Let us now consider the situation in the classical case. For one-particle systems, mixed states are identified with probability distributions in the space of densities $\operatorname{Den}(\Bbb{R}^6)$, where we have used $T^*\Bbb{R}^3\simeq\Bbb{R}^6$. The dynamics of a classical probability distribution is then given by the classical Liouville equation
\beq\label{clLiou}
\partial_t f=\{H,f\}
\,,
\eeq
where $\{\cdot,\cdot\}$ denotes the canonical Poisson bracket on $\Bbb{R}^6$ and $H$ denotes the classical Hamiltonian function. Similarly to the quantum case, if we denote by ${\cal S}_C\subset\operatorname{Den}(\Bbb{R}^6)$ the convex subset of positive-definite (probability) distributions in the set of phase-space densities, its extreme points define the pure states. In the classical setting, the latter coincide with point measures of the type 
\beq\label{SingPart}
f({\bf q},{\bf p},t)=\delta({\bf q}-\bar{\bf q}(t))\delta({\bf p}-\bar{\bf p}(t)),
\eeq
so that the Liouville equation \eqref{clLiou} returns Hamilton's equations
\[
\dot{\bar{\mathbf{q}}}=\partial_{\bar{\bf p}}H(\bar{\mathbf{q}},\bar{\mathbf{p}})\,,\qquad\qquad
\dot{\bar{\mathbf{p}}}=-\partial_{\bar{\bf q}}H(\bar{\mathbf{q}},\bar{\mathbf{p}})
\,,
\]
for the motion of the single particle in the system. For further details, see \cite{ChernoffMarsden} and \cite{Shirokov}. 

Again, the mapping
\beq\label{Klimontovichmomap}
(\bar{\mathbf{q}},\bar{\mathbf{p}})\mapsto \delta({\bf q}-\bar{\bf q})\delta({\bf p}-\bar{\bf p})
\eeq
is the general case of an equivariant momentum map $T^*\Bbb{R}^3\to\operatorname{Den}(\Bbb{R}^6)$ which first appeared in \cite{MaWe83} and was later studied in \cite{GBTrVi,HoTr09,GBVi}. The geometry underlying this momentum map is somewhat involved and goes back to van Hove's thesis \cite{VanHove} in 1951. As mentioned, the map \eqref{Klimontovichmomap} takes phase-space in the space $\operatorname{Den}(\Bbb{R}^6)$ of densities, which is identified with the dual of phase-space functions in $C^\infty(\Bbb{R}^6)$. In turn, the latter space is a Lie algebra under the canonical Poisson bracket $\{\cdot,\cdot\}$ and this Lie algebra integrates to an infinite dimensional group that was studied in detail by van Hove. Previously called ``contact transformations'' by Dirac \cite{Dirac1,Dirac2} (following Lie \cite{Lie}), the elements of this group were later named \emph{strict contact transformations} \cite{Gray} as they apply only to autonomous Hamiltonian systems (i.e. with a time-independent Hamiltonian). In this context, the {\it prequantum bundle} $T^*\Bbb{R}^3\times S^1\simeq\Bbb{R}^6\times S^1$ is a contact manifold with contact 1-form ${\cal A}+\de \tau$, where ${\cal A}=-{\bf p}\cdot\de {\bf q}$ is the canonical one form on $T^*\Bbb{R}^3\simeq\Bbb{R}^6$. 
The contact form identifies a connection on the prequantum bundle and this connection has the local form 
${\cal A}$. In this setting, strict contact transformations are given by connection-preserving bundle automorphisms, that is
\begin{equation}\label{stricts}
\operatorname{Aut}_{\cal A}(\Bbb{R}^6\times S^1)=
\left\{(\eta,e^{i\varphi})\in\operatorname{Diff}(\Bbb{R}^6)\,\circledS\, \mathcal{F}(\Bbb{R}^6, S^1)\ \Big|\ \eta^*\mathcal{A}+\de\varphi=\mathcal{A} \right\}
.
\end{equation}
Here, the symbol $\circledS$ denotes the semidirect product, $*$ denotes pullback, and $\de$ is the exterior differential on $\Bbb{R}^6$ (so that $\de\varphi=\nabla\varphi$). It is clear that $\eta^*\de{\cal A}=\de{\cal A}$ and also $\varphi=\theta+\int_{\boldsymbol{0}}^\bz({\cal A}-\eta^*{\cal A})$, where $\theta=\varphi(\boldsymbol{0})$ and the line integral is computed along an arbitrary curve connecting the origin to the point $\bz$. Notice that we shall use the notation $\mathcal{F}(M, N)$ to indicate the space of mappings from the manifold $M$ to the manifold $N$.

A more convenient setting for dealing with these transformations is provided by  central extensions. Indeed, the group \eqref{stricts} of strict contact transformations is isomorphic to a central extension of canonical transformations (that is, Hamiltonian diffeomorphisms, denoted by $\operatorname{Diff}_\textrm{\tiny Ham}(\Bbb{R}^6)$) by the circle $S^1$ and its multiplication rule reads as follows \cite{GBTr12,GBTrVi,IsLoMi2006}:
\beq
(\eta_1,e^{i\kappa_1})(\eta_2,e^{i\kappa_2})=\bigg(\eta_1\circ\eta_2,\,\exp\bigg(i\kappa_1+i\kappa_2+i\int_{\boldsymbol{0}}^{\eta_2(\boldsymbol{0})}(\eta_1^*\mathcal{A}-\mathcal{A})\bigg)\bigg)
\,,
\eeq
where $(\eta_j,\kappa_j)\in  \operatorname{Diff}_\textrm{\tiny Ham}(\Bbb{R}^6)\times S^1$ and $\circ$ denotes composition. Notice that here the Cartesian product symbol $\times$ stands as an abuse of notation because the group $\operatorname{Diff}_\textrm{\tiny Ham}(\Bbb{R}^6)\times S^1$ is actually a central extension (not a direct product). This group extension possesses the natural left action on $\Bbb{R}^6$, given by $({\bf p}, {\bf q})\mapsto\eta({\bf p}, {\bf q})$, and whose infinitesimal generator is $({\bf p}, {\bf q})\mapsto\mathbf{X}_H({\bf p}, {\bf q})$ (here, $\mathbf{X}_H$ denotes the Hamiltonian vector field generating the canonical transformation $\eta$). Notice that, in this case, the group action is not linear and thus relations such as the first equality in \eqref{momapformula1} cannot be used. However,  \eqref{Klimontovichmomap} is easily seen to be a momentum map \cite{GBTrVi,HoTr09} upon verifying the Poisson bracket formula
\beq
\left\{F,\int \!J(\bar{\bf q},\bar{\bf p})H({\bf q},{\bf p})\,\de^3{q}\,\de^3{p} \right\}=\mathbf{X}_H[F]=-\{H,F\}
\,,\qquad
\qquad
\forall F\in C^\infty (\Bbb{R}^6)
\eeq
for nonlinear symplectic actions. Indeed, this relation is verified immediately by setting $J(\bar{\bf q},\bar{\bf p})=\delta({\bf q}-\bar{\bf q})\delta({\bf p}-\bar{\bf p})$. In addition, this momentum map is manifestly equivariant since $\det\nabla\eta=1$ and thus $(J\circ\eta)(\bar{\bf q},\bar{\bf p})=\eta^*J(\bar{\bf q},\bar{\bf p})$.

The above correspondence between pure and mixed states in quantum and classical mechanics is the   point of departure for this paper, which shows how the fundamental momentum maps reported above can be immediately generalized to yield momentum map pair structures in different contexts.

\section{Mixed states and momentum maps}

The setting outlined in the previous section can be immediately generalized to what is usually called \emph{mixtures} in the context of quantum mechanical states. The same concept also applies to classical mechanics within the so-called \emph{Klimontovich method} of kinetic theory \cite{Kl1967}. Before beginning the discussion, it is important to remark that here we continue to consider a one-particle system in both the quantum and the classical setting.

\subsection{Quantum mixtures as momentum maps}
In quantum mechanics, mixed states are often expressed in terms of mixtures of (non-orthogonal) pure states as follows:
\beq\label{mixture}
\rho = \sum_{k}^N w_k\psi_k\psi^\dagger_k
\,,
\eeq
where $\psi_k\in\mathscr{H}$. Here, the number $N$ has nothing to do with the number of particles in the system, since here we only deal with one particle. In standard textbooks \cite{Pauli,vonNeumann}, the relation \eqref{mixture} is usually interpreted as  a mixture of pure states for the one-particle system, where $w_i$ indicates the probability of the $k-$th pure state $\psi_i$ in the mixture. For consistency, one also requires that $\|\psi_k\|^2=1$, so that $\operatorname{Tr}\rho=\sum_k w_k = 1$. In analogy to the case of pure states, see \eqref{PStates}, this normalization will be recovered here as an initial condition that is preserved by unitary dynamics.

In order to unfold the momentum map features of \eqref{mixture}, let us define the following symplectic form on the Cartesian product $\mathscr{H}\times\dots\times\mathscr{H}$:
\beq
\Omega(\{\psi_k^{(1)}\},\{\psi_k^{(2)}\})=2\hbar\sum_k^Nw_i\operatorname{Im}\langle\psi_k^{(1)}|\psi_k^{(2)}\rangle
\,,\qquad\qquad
\{\psi_k^{(1)}\},\{\psi_k^{(2)}\}\in\mathscr{H}\times\dots\times\mathscr{H}
\eeq
Then, it is immediate to see that the quantity $-i\hbar\rho=-i\hbar\sum_{k} w_k\psi_k\psi_k^\dagger$ identifies an equivariant momentum map of the type
\[
\mathscr{H}\times\dots\times\mathscr{H}\to\mathfrak{u}(\mathscr{H})^*
,
\]
for the natural right action $\{\psi_k\}\mapsto\{U\psi_k\}$ of unitary operators $U\in {\cal U}(\mathscr{H})$. 
In turn, this momentum map leads to a sequence of Schr\"odinger equations for each pure state $\psi_k$, that is
\beq
i\hbar\partial_t\psi_k=H\psi_k
\,.
\eeq
This can be verified by simply replacing \eqref{mixture} in the quantum Liouville equation \eqref{QLiouville} for $\rho$. Here, each wavefunction $\psi_k$ evolves under the unitary propagator $U(t)=\exp(-i Ht/\hbar)$, so that $\psi_k(t)=U(t)\psi_k^{(0)}$ and the normalization condition $\|\psi_k\|^2=1$ is preserved in time. Since the quantum Liouville equation \eqref{QLiouville} also preserves $\operatorname{Tr}\rho=1$, this leads immediately to $\sum_kw_k=1$. Similar considerations will also apply to other cases throughout this paper.

The above picture can be further extended upon replacing the sequence $\{\psi_k\}_{k=1\dots N}$ by a continuous family of wavefunctions $\psi_k(r)$ parameterized by a set of coordinates $r\in\Bbb{R}^n$. In this setting, the normalization condition becomes $\|\psi(r)\|^2=1$ and the weights $w_k$ are replaced by the measure $w(r)\in\operatorname{Den}(\Bbb{R}^n)$, so that $(\Bbb{R}^n,w)$ becomes a volume vector space and  \eqref{mixture} generalizes to
\beq\label{genmixture}
\rho=\int\!w(r)\,\psi(r)\psi^\dagger(r)\,\de^n r
\,.
\eeq
For example,  this type of expression emerges in dynamical models for nonadiabatic molecular dynamics \cite{FoHoTr18,IsLoMi2006}, where it determines the density operator  for the electronic dynamics. As it is shown below, this expression determines the left leg of a dual pair of momentum maps underlying quantum dynamics. The momentum map character of the quantity $-i\hbar\int\!w(r)\,\psi(r)\psi^\dagger(r)\,\de^n r$ is easy to see. If we denote by $\mathcal{F}(\Bbb{R}^n,\mathscr{H})$ the set of wavefunctions in $\mathscr{H}$ that are parameterized by $r\in \Bbb{R}^n$, it suffices to construct the symplectic form
\beq\label{BigSympQForm}
\Omega\big(\psi^{(1)},\psi^{(2)}\big)=2\hbar\operatorname{Im}\!\int \!w(r)\,\big\langle\psi^{(1)}(r)\big|\psi^{(2)}(r)\big\rangle\,\de^n r
\eeq
to observe that the generalized mixture \eqref{genmixture} identifies a momentum map for the natural right action $\psi(r)\mapsto U\psi(r)$ of unitary operators $U\in {\cal U}(\mathscr{H})$. 
We remark that the symplectic form \eqref{BigSympQForm} is strictly related to a class of symplectic forms previously appeared in \cite{GBTrVi,GBVi}; let $S$ be a compact orientable manifold with volume form $\mu_S$ and let $(M, \omega)$ be an exact symplectic manifold. One can endow the manifold $\mathcal{F}(S,M)$ of smooth functions $S\to M$ with the symplectic form
\begin{align*}
 \bar{\omega}(f) (u_f,v_f)=\int_S \omega(f(x))(u_f(x),v_f(x))\mu_S. 
\end{align*}
In our case,  $M=\mathscr{H}$ and the symplectic form above recovers \eqref{BigSympQForm} upon replacing $S$ by $\mathbb{R}^n$ and by setting $\mu_S=w(r)\,\de^n r$. Since $\mathbb{R}^n$ is not a compact manifold, special care must be taken in ensuring that the integral in \eqref{BigSympQForm} converges; however, here we proceed formally without
 dealing with these important issues.

It is obvious that the above picture can be generalized further to consider a sequence of volume forms $\{w_k(r) \de^nr\}$ on $\Bbb{R}^n$ so that the Cartesian product $\mathcal{F}(\Bbb{R}^n,\mathscr{H})\times\dots\times \mathcal{F}(\Bbb{R}^n,\mathscr{H})$ can be endowed with the symplectic form
\beq\label{BigSympQForm2}
\Omega\big((\{\psi_k^{(1)}(r)\},\{\psi_k^{(2)}(r)\})\big)=2\hbar\operatorname{Im}\!\left[\sum_{k=1}^N\int \! w_k(r)\,\big\langle\psi_k^{(1)}(r)\big|\psi_k^{(2)}(r)\big\rangle\,\de^n r\right]
,
\eeq
thereby leading to an equivariant momentum map associated to the density matrix
\beq\label{genmixture2}
\rho=\sum_{k=1}^N\int\!w_k(r)\,\psi_k(r)\psi_k^\dagger(r)\,\de^n r
\,,
\eeq
which generalizes the previous expressions.

\subsection{Klimontovich approach to classical mechanics\label{sec:Klim1}}
The arguments in the previous section transfer immediately to the classical setting. For example, the momentum map \eqref{SingPart} immediately extends to the sampling distribution
\beq
f({\bf q},{\bf p})=\sum_{k}^Nw_{k\,}\delta({\bf q}-\bar{\bf q}_k)\delta({\bf p}-\bar{\bf p}_k)\,,
\label{multiKlim}
\eeq
where again $\int \! f\,\de^3q\,\de^3p = \sum_k w_k = 1$. Analogously to the quantum case, one defines the following symplectic form on $\Bbb{R}^{6N}$: 
\beq
\Omega\big(\{(\bar{\bf q}_k,\bar{\bf p}_k)\}\big)=\sum_{k}^Nw_{k\,}\de\bar{\bf q}_k\wedge\de \bar{\bf q}_k
\label{ncsymp}
\eeq
and verifies that the natural action $\{(\bar{\bf q}_k,\bar{\bf p}_k)\}\mapsto\{\eta(\bar{\bf q}_k,\bar{\bf p}_k)\}$ of the group $\operatorname{Diff}_\textrm{\tiny Ham}(\Bbb{R}^6)\times S^1$ on $\Bbb{R}^{6N}$ determines a momentum map which coincides with \eqref{multiKlim}. Again, notice that the number $N$ has generally nothing to do with the number of particles: indeed, in our setting the system under consideration comprises only one particle whose probability density is given by the distribution $f({\bf q},{\bf p})$. As  mentioned above, the expression \eqref{multiKlim} can be interpreted as a \emph{classical mixture} in terms of a standard sampling process in statistics. Nevertheless, we observe that replacing \eqref{multiKlim} in the classical Liouville equation \eqref{clLiou} does return a multi-body system obeying canonical equations
\[
\dot{\bar{\mathbf{q}}}_k=\frac1{w_k}\frac{\partial H}{\partial {\bar{\bf p}_k}}
\,,\qquad\qquad
\dot{\bar{\mathbf{p}}}_k=-\frac1{w_k}\frac{\partial H}{\partial {\bar{\bf q}_k}}
\,,
\]
as it is prescribed by the collectivization theorem of Guillemin and Sternberg \cite{Guillemin}. Notice that these Hamiltonian equations are not strictly canonical since they are associated to the symplectic form \eqref{ncsymp}, which itself is not exactly canonical. The same argument holds, for example, in point vortex motion \cite{MaWe83}.

In previous work \cite{GBTrVi,HoTr09}, the author considered the following extension of the above construction. Upon replacing the sequence $\{(\bar{\bf q}_k,\bar{\bf p}_k)\}_{k=1\dots N}$ by a continuous family of points $(\bar{\bf q}(r),\bar{\bf p}(r))$ parameterized by a set of coordinates $r\in\Bbb{R}^n$, one can construct the  distribution
\beq
f({\bf q},{\bf p})=\int\!w(r)\,\delta({\bf q}-\bar{\bf q}(r))\delta({\bf p}-\bar{\bf p}(r))\,\de^n r\,,
\label{multiKlim2}
\eeq
where $w(r)\in\operatorname{Den}(\Bbb{R}^n)$.
Once again, one can construct a symplectic form \cite{GBTrVi,GBVi} on $\mathcal{F}(\Bbb{R}^n,\Bbb{R}^6)$
\[
\Omega(\mathbf{X},\mathbf{Y})=\int w(r) X^a(r)\mathbb{J}_{ab}Y^b(r)\,\de^nr
\,,
\qquad\qquad
\forall\,\mathbf{X}, \mathbf{Y}\in \mathcal{F}(\Bbb{R}^n,\Bbb{R}^6)
\]
where $\mathbb{J}_{ab}$ is the canonical symplectic form. Then, the relation \eqref{multiKlim2} identifies an equivariant momentum map for the natural action $\big(\bar{\bf q}(r),\bar{\bf p}(r)\big)\mapsto\eta\big(\bar{\bf q}(r),\bar{\bf p}(r)\big)$ of the group $\operatorname{Diff}_\textrm{\tiny Ham}(\Bbb{R}^6)\times S^1$ on the symplectic space $\mathcal{F}(\Bbb{R}^n,\Bbb{R}^{6})$.

To continue the analogy with the previous section, we can also generalize further the construction above and consider a sequence of volume forms $\{w_k(r) \de^nr\}$ on $\Bbb{R}^n$ so that the Cartesian product $\mathcal{F}(\Bbb{R}^n,\Bbb{R}^{6})\times\dots\times \mathcal{F}(\Bbb{R}^n,\Bbb{R}^{6})$ can be endowed with a suitable symplectic form, thereby leading to the momentum map
\beq
f({\bf q},{\bf p})=\sum_{k=1}^N\int\!w_k(r)\,\delta({\bf q}-\bar{\bf q}_k(r))\delta({\bf p}-\bar{\bf p}_k(r))\,\de^n r\,.
\label{multiKlim3}
\eeq
For example, expressions of this type were considered in \cite{HoTr09}, where they were also related to the singular solutions emerging in certain types of hydrodynamic PDEs, known as \emph{EPDiff equations} \cite{HoMa2005}.

\section{Right actions and diffeomorphisms}

In the previous sections, all momentum maps appearing in quantum and classical mixed states were associated to specific left actions of  ${\cal U}(\mathscr{H})$ and $\operatorname{Diff}_\textrm{\tiny Ham}(\Bbb{R}^6)\times S^1$, respectively. In the particular case when the representation spaces are $\mathcal{F}(\Bbb{R}^n,\mathscr{H})$ and $\mathcal{F}(\Bbb{R}^n,\Bbb{R}^{6})$, with $\Bbb{R}^n$ carrying the volume form $w=w(r)\,\de^n r$, additional momentum maps can be constructed by considering the pullback action of the group  $\operatorname{Diff}_\textrm{\tiny vol}(\Bbb{R}^n)$ of volume-preserving diffeomorphisms of $\Bbb{R}^n$. In the case of classical mechanics, this fact led Marsden and Weinstein \cite{MaWe83} to construct a dual pair of momentum maps underlying planar incompressible fluid flows. As reported also in the sections below, this construction has recently been developed further in \cite{GBTrVi,GBVi}, while the application to Liouville-type (Vlasov) equations was presented in \cite{HoTr09}.  The following section shows that an analogue construction also underlies quantum mixed states.

\subsection{The Berry curvature as a momentum map}

As mentioned above, the space $\mathcal{F}(\Bbb{R}^n,\mathscr{H})$ comprising the wavefunctions $\psi(r)$ (also known as \emph{electronic wavefunctions} in molecular dynamics \cite{Baer,Marx}) carries two different representations. On one hand, the group ${\cal U}(\mathscr{H})$ of unitary operators acts from the left, thereby generating the momentum map associated to \eqref{genmixture}. On the other hand, the group $\operatorname{Diff}_\textrm{\tiny vol}(\Bbb{R}^n)$ of volume-preserving diffeomorphisms of $\Bbb{R}^n$ carries a right (linear) action given by the pullback operation. In turn, this momentum map is given as
\beq\label{Berrymomap}
\psi(r)\mapsto\de \big\langle\psi(r)\big|-i\hbar\de\psi(r)\big\rangle=\de A(r)=:B(r)
\,.
\eeq
Here $\de$ is the differential on $\Bbb{R}^n$ and the one-form
\beq\label{Berryconn}
A=\langle\psi|-i\hbar\de\psi\rangle
\eeq
is the celebrated Berry connection \cite{berry1984quantal} on the circle bundle $\Bbb{R}^n\times S^1$, so that the equivariant momentum map for the pullback representation of $\operatorname{Diff}_\textrm{\tiny vol}(\Bbb{R}^n)$ on $\mathcal{F}(\Bbb{R}^n,\mathscr{H})$ is given by the Berry curvature $B=\de A$. The proof that the mapping \eqref{Berrymomap} is a momentum map is a direct verification upon applying the formula 
\beq\label{momapformula}
\langle J(\psi),\xi\rangle=-\frac12\,\Omega(\xi(\psi),\psi)
\,,
\eeq
where the symplectic form $\Omega$ is given in \eqref{BigSympQForm} and the minus sign is now due to the fact that we are dealing with a right action (different sign conventions for left/right actions appear in the literature). Here, the Lie algebra element $\xi$ is given by a volume-preseving vector field acting on $\psi(r)$ by Lie derivative, that is $\psi\mapsto \imath_\xi\de\psi$, where $\imath_\xi$ denotes the insertion of a vector field into a one-form. Since $\xi$ is such that $\operatorname{div}(w\xi)=0$, then 
\beq\label{incomvectflds}
\xi^\flat=w^{-1}\delta \gamma
\,,
 \eeq
where $\flat$ is the index lowering (flat) operator, $\delta$ denotes the co-differential \cite{AbMaRa} and the two-form $\gamma\in\Lambda^2(\Bbb{R}^n)$ is usually known in fluid dynamics as the \emph{stream-function}. For more details, see \cite{GBVi,MaWe83}. Since we are in $\Bbb{R}^n$, we can drop the flat symbol by using the Euclidean metric. In the general case, the relation \eqref{incomvectflds} defines a Lie algebra isomorphism between the space $\mathfrak{X}_\textrm{\tiny $\rm vol$}(\Bbb{R}^n)$ of incompressible vector fields and the space $\Lambda^2(\Bbb{R}^n)/\Bbb{R}$ of two-forms modulo real numbers. At this point, it suffices to expand the right hand side of \eqref{momapformula} to get
\beq
-\frac12\,\Omega(\xi(\psi),\psi)=-\hbar\int   \langle\psi|i\de\psi\rangle\wedge*\delta\gamma=\langle  A,\delta\gamma\rangle
=\langle \de A,\gamma\rangle
\,,
\eeq
where we have used the Hodge star operator $*:\Lambda^{k}(\Bbb{R}^n)\to\Lambda^{n-k}(\Bbb{R}^n)$ and integration by parts under the Hodge pairing. Then, since the space ${\xi\in\mathfrak{X}_\textrm{\tiny $\rm vol$}(\Bbb{R}^n)}$ is identified with $\gamma\in \Lambda^2(\Bbb{R}^n)/\Bbb{R}$, the dual space $\mathfrak{X}_\textrm{\tiny vol}(\Bbb{R}^n)^*$ of incompressible vector fields can be identified with the space $\de\Lambda^1(\Bbb{R}^n)$ of exact two-forms and thus the relation \eqref{momapformula} returns the momentum map
\beq\label{Berry1}
J(\psi)=\de A=\hbar\operatorname{Im}(\de\psi^\dagger\wedge\de\psi)\in \Lambda^2(\Bbb{R}^n)\simeq\mathfrak{X}_\textrm{\tiny vol}(\Bbb{R}^n)^*
\,,
\eeq
indeed coinciding with the Berry curvature $B:=\de A$.  Here, it may be useful to remark that this picture may also be extended to the generalized case associated to the density matrix expression \eqref{genmixture2} upon considering the direct product group $\operatorname{Diff}_\textrm{\tiny vol}^{(1)}(\Bbb{R}^n)\times\dots\times\operatorname{Diff}_\textrm{\tiny vol}^{(N)}(\Bbb{R}^n)$, with $\operatorname{Diff}_\textrm{\tiny vol}^{(k)}(\Bbb{R}^n)=\{\eta\in\operatorname{Diff}(\Bbb{R}^n)\ | \ \eta^*w_k=w_k\}$.

To summarize, the space $\mathcal{F}(\Bbb{R}^n,\mathscr{H})$ of parameterized (electronic) wavefunctions is a representation space for two different groups, that is ${\cal U}(\mathscr{H})$ (acting from the left) and $\operatorname{Diff}_\textrm{\tiny vol}(\Bbb{R}^n)$ (acting from the right). Both these groups carry Hamiltonian actions producing momentum maps summarized as follows:
\begin{equation}\label{QuantuDP}
\mathfrak{u}(\mathscr{H})^*\longleftarrow \mathcal{F}(\Bbb{R}^n,\mathscr{H})\longrightarrow \mathfrak{X}_\textrm{\tiny vol}(\Bbb{R}^n)^*
\,,
\end{equation}
where the left leg  corresponds to the relation \eqref{genmixture} and the right leg is given by \eqref{Berrymomap}. 

Special cases of similar constructions are provided by \emph{dual pairs}, in which the kernels of the two momentum maps enjoy a symplectic orthogonality condition \cite{GBTrVi,GBVi,Weinstein}. For example, a different pair of momentum maps in the context of quantum mixed states was found to be a dual pair in \cite{Montgomery91}. In this case, $\psi\in\mathscr{H}=\mathscr{H}^{(1)}\otimes\mathscr{H}^{(2)}$ and the partial traces $\rho_2=\operatorname{Tr}_{\mathscr{H}^{(1)}}\psi\psi^\dagger$ and $\rho_1=\operatorname{Tr}_{\mathscr{H}^{(2)}}\psi\psi^\dagger$ were found to identify momentum maps for the natural actions of $U(\mathscr{H}^{(2)})$ and $U(\mathscr{H}^{(1)})$, respectively. Then, the momentum map pair $\mathfrak{u}(\mathscr{H}^{(2)})^*\leftarrow \mathscr{H}\rightarrow \mathfrak{u}(\mathscr{H}^{(1)})^*
$ was found to be a dual pair.

In the classical case, a  construction similar to \eqref{QuantuDP} also leads to a dual pair of momentum maps \cite{GBTrVi,GBVi,MaWe83} and this is reported in the following section.

\subsection{The dual pair of classical mechanics}
As discussed above, the group of volume-preserving diffeomorphisms has a natural pullback action on the space of parameterized wavefunctions. Likewise, in the classical setting the same group acts by pullback on the space $\mathcal{F}(\Bbb{R}^n,\Bbb{R}^{6})$ of generalized coordinates $\big(\bar{\bf q}(r),\bar{\bf p}(r)\big)$ from Section \ref{sec:Klim1}. Since this representation is also Hamiltonian, it leads to an equivariant momentum map that is expressed as \cite{GBTrVi,GBVi,HoTr09}
\begin{equation}\label{ClassicalBconn}
\big(\bar{\bf q}(r),\bar{\bf p}(r)\big)\mapsto-\de\big(\bar{p}_a(r)\de\bar{q}^a(r)\big)=\de\bar{q}^a(r)\wedge\de\bar{p}_a(r)\in 
\de\Lambda^1(\Bbb{R}^n)\simeq\mathfrak{X}_\textrm{\tiny vol}(\Bbb{R}^n)^*
\,,
\end{equation}
which is the immediate classical analogue of the Berry curvature from the previous section. Then, we are left with a similar picture to that found in the quantum case, which may be summarized as follows:
\begin{equation}\label{KlimDualPair}
 \mathfrak{X}_\textrm{\tiny Ham}(\Bbb{R}^6)^*\times\Bbb{R}\longleftarrow \mathcal{F}(\Bbb{R}^n,\Bbb{R}^6)\longrightarrow \mathfrak{X}_\textrm{\tiny vol}(\Bbb{R}^n)^*
\,.
\end{equation}
Here,  the Lie algebra $ \mathfrak{X}_\textrm{\tiny Ham}(\Bbb{R}^6)\times S^1$ of the group $\operatorname{Diff}_\textrm{\tiny Ham}(\Bbb{R}^6)\times S^1$ can be identified with the Poisson algebra $C^\infty(\Bbb{R}^6)$ via the isomorphism \cite{GBTr12}
\[
\big(\mathbf{X}_H(\bq,\bp),\gamma\big)
\mapsto
H(\bq,\bp)-H(\mathbf{0},\mathbf{0})+\gamma
\,,
\] 
so that the dual space $ \mathfrak{X}_\textrm{\tiny Ham}(\Bbb{R}^6)^*\times\Bbb{R}$ can be replaced by the space of densities $\operatorname{Den}(\Bbb{R}^6)$.
 For further details about this identification and other features of the group $\operatorname{Diff}_\textrm{\tiny Ham}(\Bbb{R}^6)\times S^1$, we refer the reader to \cite{GBTr12,IsLoMi2006}. 
In the momentum map pair \eqref{KlimDualPair}, the left leg is given by the generalized Klimontovich solution \eqref{multiKlim2}, while the right leg is given as in \eqref{ClassicalBconn}. Interestingly enough, in the classical case these momentum maps  are known to produce a dual pair structure, as discussed in \cite{GBTrVi,GBVi,HoTr09}. 

We conclude this section by emphasizing its main result: analogous momentum map pairs occur in \emph{both} quantum and classical mechanics. While the left leg reproduces quantum mixtures and Klimontovich solutions (respectively, in the quantum and the classical case), the right leg yields the Berry curvature and its classical analogue. The next section will apply the momentum maps occurring in the quantum case to certain models currently used in molecular dynamics simulations.

\section{Momentum maps in quantum molecular dynamics}

This section unfolds how the above momentum maps for quantum mixtures appear in quantum chemistry, with special focus on molecular dynamics models. In this context, one wants to solve the Schr\"odinger equation for an ensemble of particles comprising a different number of nuclei and electrons. Given the computational costs of full quantum simulations, different strategies have been designed over almost a century to approximate the nuclei as classical particles while treating the electrons in a full quantum setting.

\subsection{Born-Oppenheimer approximation and electron mixtures}
In quantum molecular dynamics, the most celebrated model is the \emph{Born-Oppenheimer approximation} \cite{born1927quantentheorie}. This is based on the following decomposition for the molecular wavefunction:
\beq\label{BO}
\Psi(\br,\bx,t)=\chi(\br,t)\psi(\bx;\br)
\,,
\eeq
which is then replaced in the multi-body Schr\"odinger equation.
Here, the treatment has been simplified to consider only one electron (coordinates $\bx$) and one nucleus (coordinates $\br$). While $\chi(\br,t)$ is a genuine wavefunction, $\psi(\bx;\br)$ is considered as an $\br-$dependent wavefunction with respect to $\bx$. As such, one has the so-called \emph{partial normalization condition} (PNC)
\beq\label{PNC}
\|\psi(\br)\|^2=\int\!|\psi(\bx;\br)|^2\,\de x=1
\,.
\eeq
In the context of molecular dynamics, the parameterized wavefunction $\psi(\bx;\br)$ is time-independent (\emph{adiabatic approximation}) and is given as the fundamental eigenfunction of a specific Hamiltonian operator. Without introducing unnecessary details, it suffices to say that certain approximations are then adopted to solve the dynamics of $\chi(\br,t)$ numerically. 

It is important to remark that, while $\chi$ and $\psi$ are often referred to as \emph{nuclear} and \emph{electronic} wavefunction, respectively, these terms do not correspond to genuine pure states for the nucleus and for the electron. Indeed, as already noticed in \cite{FoHoTr18,IsLoMi2006} the molecular density operator 
\beq
\rho(\br,\bx,\br',\bx')=\chi(\br)\psi(\bx;\br)\chi^*(\br')\psi^*(\bx';\br')
\eeq
yields the following expressions for the nuclear and electronic density matrices, respectively:
\beq\label{nucelecdensities}
\rho_n(\br,\br')=\chi(\br)\chi^*(\br')\int\!\psi(\bx;\br)\psi^*(\bx;\br')\,\de x
\,,\qquad\quad
\rho_e(\br,\br')=\int\!|\chi(\br)|^{2\,}\psi(\bx;\br)\psi^*(\bx';\br)\,\de r
\,.
\eeq
Here, the explicit time dependence has been dropped for convenience. Since neither of these two operators is a projection, one concludes that neither the nucleus nor the electron are in a pure state and thus the word `wavefunction' lacks physical sense in this context. Hence, both the nucleus and the electron are in a mixed quantum state. In particular, the electronic state (second expression above) is represented exactly by  the momentum map \eqref{genmixture}. This shows that the idea of a  generalized mixture emerges naturally in molecular chemistry problems.

Nowadays, the Born-Oppenheimer approximation is often replaced by the adoption of more sophisticated methods in order to capture more dynamical features of the electronic motion. Indeed, from the second of \eqref{nucelecdensities}, we notice that the dynamics of the electron density is entirely slaved to that of the wavefunction $\chi$, which in turn is often approximated by semiclassical methods. More complete models are then necessary in order to capture nonadiabatic effects; that is, to overcome the adiabatic approximation.

\subsection{Exact factorization and the Berry curvature}

Over the last decade, a model  due to Gross and collaborators \cite{abedi2010exact,abedi2012correlated} has been receiving increasing attention, although its roots are traced back to the works of von Neumann \cite{vonNeumann} and, in later years, of Hunter \cite{Hunter}. In essence, the parameterized wavefunction is promoted to be time-dependent, so that the Born-Oppenheimer approximation \eqref{BO} is replaced by
\beq
\Psi(\br,\bx,t)=\chi(\br,t)\psi(\bx,t;\br)
\,,
\eeq
along with the PNC \eqref{PNC}, which now becomes $\|\psi(\br,t)\|^2=\int |\psi(\bx,t;\br)|^2\,\de x=1$. The dynamical model resulting from the above solution ansatz for the two-body Schr\"odinger equation is quite involved, although very rich in geometric content as recently presented in \cite{FoHoTr18}, where analogies with complex fluid models were also disclosed.  In analogy to the previous sections, here we simplify the treatment by restricting to a finite-dimensional electronic Hilbert space, so that $\psi(\br)\in\Bbb{C}^n$.

A crucial ingredient emerging in the exact factorization model is the dynamical Berry connection \eqref{Berryconn}. Indeed, as outlined in \cite{agostini2015exact,FoHoTr18}, this quantity generates a Maxwell-like field thereby producing Lorentz forces in the equations of motion. Thus, the Berry curvature \eqref{Berry1} (here, $n=3$)
\[
\bB(\br,t)=\hbar\operatorname{Im}\int\!\nabla\psi(\bx,t;\br)^*\times\nabla\psi(\bx,t;\br)\,\de^3x
\]
plays an essential role in exact factorization dynamics. This is another manifestation of the emergence of momentum maps in molecular chemistry problems: the exact factorization model comprises the dynamics of \emph{both}  mappings \eqref{genmixture} (or, equivalently, the second in \eqref{nucelecdensities}) and \eqref{Berrymomap} in the momentum map pair \eqref{QuantuDP}. 

It may be important to remark that the presence of a non-zero electric-like field $\bE$ (which in this case depends on both wavefunctions $\chi$ and $\psi)$ leads to the Faraday-like equation \cite{FoHoTr18}
\beq
\partial_t\bB=-\nabla\times\bE
\,,
\eeq
so that
\beq
\frac{\de}{\de t}\oiint \bB\cdot\de\mathbf{S}=-\oint\bE\cdot\de\mathbf{x}
\neq0
\,.
\eeq
The integral of the Berry curvature over a closed surface is then related to topological singularities that form in terms of multi-valued expressions of the phase of $\psi$. In the context of the Born-Oppenheimer approximation, these singularities are related to the so called \emph{conical intersections} between energy surfaces \cite{Baer, Marx}, although this aspect will not be covered in this paper. In the case of the exact factorization model, one is left with a picture in which phase singularities may be created by the dynamics and their evolution is an important aspect of the model (unlike the Born-Oppenheimer case, where singularities are fixed in time). The fact that topological singularities are given by the right leg of the momentum map pair \eqref{QuantuDP} is another manifestation of the fundamental role played by momentum maps in mechanical systems.

\section{Clebsch representations}

The preceding sections have presented several types of momentum maps which emerge in both quantum and classical dynamical models. While  these were already known in the classical setting \cite{GBTrVi,GBVi,HoTr09,MaWe83}, new momentum maps were found for the case of quantum dynamics. Generally speaking, all these momentum maps are examples of \emph{Clebsch representations} \cite{Clebsch,HoKu,MaWe83}. The latter are defined as momentum maps defined on a symplectic manifold endowed with a canonical symplectic form, which is the case for the representation spaces considered so far.

The concept of a Clebsch representation may actually lead to considering special types of solutions for certain Lie-Poisson equations. This fact was first exploited by Clebsch himself in fluid dynamics \cite{Clebsch}, while the geometric construction underlying Clebsch representations was developed much later \cite{MaWe83} in terms of momentum maps generalizing the original formulation of Clebsch canonical variables. In the cases considered before, it is clear that the Clebsch representations are provided by the right legs of the momentum map pairs in \eqref{QuantuDP} and \eqref{KlimDualPair}. 

This picture allows the discovery of other types of momentum map solutions that are defined on different representation spaces carrying a canonical group action. For example, Koopman's wavefunction description of classical dynamics \cite{Koopman} has been attracting increasing attention (see e.g. \cite{Bondar,GBTr18,RaPrUrPiSoEgMoCe}) due to its analogies to quantum mechanics. However, other types of Clebsch representations also appeared in the context of density matrix evolution. For example, in 1986 Uhlmann \cite{Uhlmann} presented an alternative representation of the density matrix in terms of the evolution of linear operators on the quantum Hilbert space. This kind of alternative representations in both quantum and classical mechanics is the subject of the next sections.

\subsection{Uhlmann's quantum density operator}

Within the context of holonomy in quantum dynamics, in 1986 Uhlmann \cite{Uhlmann} wrote the density operator in terms of an abstract linear operator $W\in L(\mathcal{V},\mathscr{H})$ from some vector space $\mathcal{V}$ (which we take again finite-dimensional)   to the quantum Hilbert space. More specifically, the density operator was written as
\beq\label{Uhlmannsrho}
\rho=WW^\dagger
\,.
\eeq
It is clear that if $\mathcal{V}$ is trivial, then $W$ reduces to a wavefunction $\psi\in\mathscr{H}$. Otherwise, the density matrix \eqref{Uhlmannsrho} does not identify a pure state unless $W^\dagger W=\boldsymbol{1}$, that is $\rho^2=\rho$. One of the purposes of this section is to show that  \eqref{Uhlmannsrho} determines a Clebsch representation $L(\mathcal{V},\mathscr{H})\to\mathfrak{u}(\mathscr{H})^*$. This proof needs only two ingredients on $L(\mathcal{V},\mathscr{H})$: a canonical symplectic form and a Hamiltonian action of ${\cal U}(\mathscr{H})$. The first is simply given by 
\beq
\omega(W_1,W_2)=2\hbar\operatorname{Im}\big[\operatorname{Tr}(W_1^\dagger W_2)\big]
\,,
\eeq
while the left action of ${\cal U}(\mathscr{H})$ is given by
\beq
W\mapsto UW
\,.
\eeq
Since the infinitesimal generator reads $W\mapsto \xi W$, with $\xi \in\mathfrak{u}(\mathscr{H})$, we compute
\beq
\frac12\omega(\xi W,W)=\hbar\operatorname{Re}\big[\operatorname{Tr}(i W^\dagger \xi W)\big]=\langle-i\hbar WW^\dagger,\xi\rangle
\,,
\eeq
thereby leading to the momentum map $W\mapsto-i\hbar\rho$. Now, since the space $L(\mathcal{V},\mathscr{H})$ is canonical, $W$ satisfies canonical Hamiltonian motion so that replacing \eqref{Uhlmannsrho} in the quantum Liouville equation yields the following Schr\"odinger-type equation on $L(\mathcal{V},\mathscr{H})$:
\beq
i\hbar \partial_tW= HW
\,.
\eeq

Notice that, in the special case $\mathcal{V}=\mathscr{H}$, the operator $W$ is a linear operator on the quantum Hilbert space $\mathscr{H}$ (that is $W\in L(\mathscr{H})$). In this particular case, the unitary group ${\cal U}(\mathscr{H})$ carries the alternative representation
\[
W\mapsto UWU^\dagger
\,,
\]
whose infinitesimal generator reads $W\mapsto [\xi,W]$.
In this particular setting, the corresponding momentum map reads
\[
W\mapsto -i\hbar[W,W^\dagger]
\,.
\]
Notice, however, that this momentum map does not produce a Clebsch representation for the density operator $\rho$, since $\operatorname{Tr}[W,W^\dagger]=0$. Still, this last construction can be adopted to provide a generalized Clebsch representation for $\rho$ that is defined on the Cartesian product $\mathscr{H}\times L(\mathscr{H})$. Indeed, upon importing the natural product symplectic form on $\mathscr{H}\times L(\mathscr{H})$, the Hamiltonian action
\[
(\psi,W)\mapsto (U\psi,UWU^\dagger)
\]
produces the momentum map $(\psi,W)\mapsto-i\hbar\rho$, with 
\[
\rho=\psi\psi^\dagger+[W,W^\dagger]
\,.
\]
Here, $\operatorname{Tr}\rho=1$ and $\rho>0$ are both preserved by the unitary evolution $\rho=U\rho_0U^\dagger$, which in turn preserves also the purity of the state since $\rho^2-\rho=U(\rho_0^2-\rho_0)U$. Then substitution of the above expression in the quantum Liouville equation yields the uncoupled equations
\[
i\hbar\partial_t\psi=H\psi
\,,\qquad\qquad
i\hbar\partial_t W=[H,W]
\,.
\]

It is not known whether this type of momentum map solutions of the quantum Liouville equation \eqref{QLiouville} may have any physical meaning. It is certainly true that the density operator in quantum mechanics is only defined up to a commutator and this observation might be used to formulate generalized theories of quantum mechanics. However, these are beyond the scope of this paper.

\subsection{Wavefunctions in classical mechanics}

In the classical setting, a Clebsch representation for the Liouville equation has been known since the early 80's \cite{Morrison} and it is essentially an immediate generalization of the Clebsch representation for the vorticity of planar incompressible flows. If $(S,D)\in T^*\mathcal{F}(\Bbb{R}^{6},S^1)$, then a Clebsch representation momentum map $T^*\mathcal{F}(\Bbb{R}^{6},S^1)\to\operatorname{Den}(\Bbb{R}^6)$ is given as
\beq\label{Clebsch1}
(S,D)\mapsto\{D,S\}
\,,
\eeq
where we recall that $\{\cdot,\cdot\}$ denotes the canonical Poisson bracket. This momentum map is associated to the cotangent lift of the natural right action of $\operatorname{Diff}_\textrm{\tiny Ham}(\Bbb{R}^6)\times S^1$ on $\mathcal{F}(\Bbb{R}^{6},S^1)$, that is given by the pullback $S\mapsto\eta^*S$ with $(\eta,\kappa)\in \operatorname{Diff}_\textrm{\tiny Ham}(\Bbb{R}^6)\times S^1$. The next section shows how this construction applies to the Koopman-von Naumann formulation of classical mechanics \cite{Koopman,VonNeumann2}. In this section, we are forced to work with infinite-dimensional Hilbert spaces and thus the discussion proceeds only formally. A more detailed presentation of these topics is currently under development \cite{GBTr19}.

\subsubsection{Koopman-von Neumann classical mechanics}
A similar structure as in \eqref{Clebsch1} can also be found by considering the symplectic Hilbert space $\mathscr{H}=L^2(\Bbb{R}^6)$ with the symplectic form in \eqref{sympform1}, that is
\beq
\omega(\psi_1,\psi_2)=2\hbar\operatorname{Im}\int\!\psi^*_1(\bz)\psi_2(\bz)\,\de^6 z
\,.
\label{KvNsympform}
\eeq
Here, we have introduced the notation $\bz=(\bq,\bp)\in\Bbb{R}^6$. In this case, the pullback action $\psi\mapsto\eta^*\psi$ of the strict contact transformation $(\eta,\kappa)\in \operatorname{Diff}_\textrm{\tiny Ham}(\Bbb{R}^6)\times S^1$ on the Hilbert space $\mathscr{H}=L^2(\Bbb{R}^6)$ gives the momentum map
\beq
\psi\mapsto i\hbar\{\psi,\psi^*\}
\,.
\eeq
Notice that the polar form $\psi=\sqrt{D}e^{iS/\hbar}$ returns exactly the expression in \eqref{Clebsch1}. 
Then, we notice that replacing the Clebsch representation 
\beq\label{Clebsch2cl}
f(\bz)=i\hbar\{\psi(\bz),\psi^*(\bz)\}
\eeq
in the Liouville equation \eqref{clLiou} yields the evolution equation for $\psi$, which can be written in the Schr\"odinger-like form
\beq\label{Koopmaneq}
i\hbar\partial_t\psi = {\sf L}_H\psi
\,,\qquad\qquad
{\sf L}_H := i\hbar\{H,\ \}
\,.
\eeq
The self-adjoint operator ${\sf L}_H$ is called the \emph{Liouvillian} and the $\psi-$equation in \eqref{Koopmaneq} is the Koopman-von Neumann (KvN) equation of classical mechanics \cite{Koopman, VonNeumann2}. However, we notice that the Clebsch representation \eqref{Clebsch2cl} is not compatible with the normalization condition $\int f =1$ and thus it is not a genuine representation of classical mechanics. However, since $|\psi|^2$ satisfies the Liouville equation, the KvN construction adopts the identification 
\beq\label{phasemomap}
f(\bz)=|\psi(\bz)|^2
\eeq
in place of \eqref{Clebsch2cl}. We note in passing that the quantity $|\psi|^2$ is itself another momentum map for the action $\psi(\bz)\mapsto e^{-i\theta(\bz)/\hbar\,}\psi(\bz)$ of local phases  $\theta(\bz)\in \mathcal{F}(\Bbb{R}^{6},S^1)$ on the Hilbert space $L^2(\Bbb{R}^6)$. 

\subsubsection{Koopman-van Hove classical mechanics}
Here, we are left with a picture in which the KvN equation is Hamiltonian with symplectic form \eqref{KvNsympform} and Hamiltonian functional $h(\psi)=i\hbar\int\!H\{\psi^*,\psi\}$. The latter differs from the physical total energy, which instead would read $\int\!H|\psi|^2$ by following the KvN prescription $f=|\psi|^2$. This apparent inconsistency, was recently overcome in \cite{GBTr18} by considering an alternative action of $\operatorname{Diff}_\textrm{\tiny Ham}(\Bbb{R}^6)\times S^1$ on classical wavefunctions $\psi\in L^2(\Bbb{R}^6)$. As reported in van Hove's thesis, this action is given by
\beq\label{KvHaction}
\psi\mapsto e^{i\hbar^{-1}\left[\kappa+\int_0^\bz(\eta^*{\mathcal{A}}-{\mathcal{A}})\right]\,}\eta^*\psi
\,,\qquad\qquad
(\eta,\kappa)\in \operatorname{Diff}_\textrm{\tiny Ham}(\Bbb{R}^6)\times S^1
\,,
\eeq
where ${\mathcal{A}}=-\bp\cdot\de\bq$ is the \emph{symplectic potential}  such that the canonical symplectic form on $\Bbb{R}^6$ is given as $\omega_\textrm{\tiny can}=\de{\mathcal{A}}$.
In turn, as shown in \cite{GBTr18}, this action produces the Clebsch representation momentum map
\begin{align}\nonumber
f=&\ |\psi|^2+\operatorname{div}\!\left[\psi^*\Bbb{J}\left({\mathcal{A}}-i\hbar\nabla\right)\psi\right]
\\
=&\ |\psi|^2+\operatorname{div}(|\psi|^2\Bbb{J}{\mathcal{A}})+i\hbar\{\psi,\psi^*\}
\,.
\label{ClebschKvH}
\end{align}
While comprising both momentum maps appearing previously in this section, this representation has the advantage that $\int f = \int |\psi|^2=1$, for a suitably normalized wavefunction. In turn, replacing \eqref{ClebschKvH} in the Liouville equation \eqref{clLiou} yields a modified version of the KvN equation  \eqref{Koopmaneq} previously appeared in \cite{Gunther,Kostant}, that is
\beq\label{KvHeq}
i\hbar\partial_t\psi = \mathcal{L}_H\psi
\,,\qquad\qquad
\mathcal{L}_H := {\sf L}_H-L
\,.
\eeq
Here,
\beq
L:={\bf X}_H\cdot\boldsymbol{\mathcal{A}}-H
\eeq
is the \emph{Lagrangian function}, as it arises from the phase term in the group action \eqref{KvHaction}. Notice that this group action produces the infinitesimal generator $i\hbar\mathcal{L}_H$, which in turn satisfies $[\mathcal{L}_H,\mathcal{L}_K]=i\hbar\mathcal{L}_{\{H,K\}}$. Partly inspired by Kirillov \cite{Kirillov}, this equation \eqref{KvHeq} has been called \emph{Koopman-van Hove (KvH) equation} and the self-adjoint operator $\mathcal{L}_H$ is called \emph{prequantum operator} in prequantization theory \cite{Hall}. As mentioned earlier, the KvH equation \eqref{KvHeq} first appeared in \cite{Gunther,Kostant}, although the relation \eqref{ClebschKvH} between the classical wavefunction $\psi(\bz)$ and the Liouville density function was discovered only recently in \cite{GBTr18}.

The main relation between the KvN and KvH constructions is that KvH reproduces the KvN equation for the modulus $D=|\psi|^2$, while it also carries the evolution for the phase. Indeed, the polar form $\psi=\sqrt{D}e^{iS/\hbar}$ yields the relations \cite{Klein}
\beq
\partial_t S = \{H,S\} + L
\,,\qquad\qquad
\partial_t D = \{H,D\} \,.
\eeq
The classical phase is then a fundamental ingredient of KvH theory, which therefore can be regarded as a completion of the KvN construction. The dynamics of the phase can be written in terms of ${\mathcal{A}}$ as follows. Using  $\pounds_{\mathbf{X}_H}{\mathcal{A}}=-\de L$ (here, $\pounds_{\mathbf{X}_H}$ denotes the Lie derivative along $\mathbf{X}_H$) leads to 
\[
(\partial_t+\pounds_{\mathbf{X}_H})(\de S+ {\mathcal{A}})=0
\,,
\]
which produces the relation $\eta^*(\de S+ {\mathcal{A}})={const.}$ Then, upon setting the constant to be ${\mathcal{A}}$ itself, we have the usual relation \cite{MaRa2013} $\de S=\eta_*{\mathcal{A}}-{\mathcal{A}}$. This is simply another manifestation of the evolution equation for the classical wavefunction
\beq
\psi=e^{i\hbar^{-1}\int_0^\bz(\eta_*{\mathcal{A}}-{\mathcal{A}})\,}\eta_*\psi_0
\eeq
(up to a global phase factor), as it emerges by formally integrating the KvH equation \eqref{KvHeq}.

The KvH construction was recently used in \cite{GBTr18} to formulate a classical-quantum wave equation for the Hamiltonian dynamics of hybrid classical-quantum systems. Such a formulation has been an open question for over 40 years, since Sudarshan's first proposal \cite{Sudarshan} of using KvN for modeling hybrid systems.  The fact that the Clebsch representation  \eqref{ClebschKvH} has finally led to a consistent Hamiltonian theory for classical-quantum dynamics is among the best successes of momentum map methods. 

\section{Conclusions}

This paper has disclosed various types of momentum maps underlying both quantum and classical dynamics. While most of them were already known in the classical setting, new momentum map features were presented for mixed quantum states and it was shown how they  emerge in dynamical models for molecular dynamics. As an example, we showed how the celebrated Berry curvature determines the right leg of a momentum map pair, whose left leg identifies the electronic density matrix in the Born-Oppenheimer approximation. 

In the second part of the paper, we showed how new momentum maps produce different representations of both the quantum density matrix and the classical probability density. Indeed, Uhlmann's density matrix was recovered as a special example of this construction and it would be interesting to know whether its possible generalizations could be of any physical significance. In the classical case, the Koopman-von Neumann construction was completed to include the dynamics of the classical phase, thereby leading to the Koopman-von Neumann theory. In the latter case, KvH theory is being currently used for designing hybrid classical-quantum models.

The momentum maps which appeared in this paper are fundamental objects in both quantum and classical mechanics, since they are produced by the actions of the most general groups determining the equations of motion in each case (e.g. the unitary group and the strict contact transformations). For example, many other momentum maps can be reproduced from those in this paper by appropriate projections arising from the action of suitable subgroups. It is expected that the momentum maps presented here will open the way to the development of geometric tools for new models in quantum physics and chemistry. An example is provided by the recent work \cite{FoHoTr18} on exact factorization models. 

Another interesting perspective involves the hydrodynamic picture of mixed states. This can be formulated by combining the Madelung transform \cite{Madelung} with the mixture momentum map underlying \eqref{mixture}. This approach could lead to interesting closure models in chemical physics, along the lines of the recent developments in \cite{FoHoTr18}. In analogy with the Koopman setting, this construction involves wavefunctions defined on the full infinite-dimensional Hilbert space space of square-integrable functions. Since this would involve introducing several aspects that were not treated in this paper, the discussion of quantum hydrodynamics for mixed states is left as a promising direction for further work in geometric quantum dynamics.

\paragraph{Acknowledgments.} I wish to express my deepest gratitude to Darryl Holm for his inspiring insight and his excitement over the years. I also thank him for his suggestion  to work in this direction, which led to considering quantum dynamics from a completely new perspective. Special thanks also go to the referees, whose keen comments and suggestions improved significantly the quality of this work. In addition, I wish to acknowledge several stimulating and enlightening conversations on these and related matters with Joshua Burby, Denys Bondar, Dorje Brody, Alex Close, Maurice de Gosson, Michael Foskett, Fran\c{c}ois Gay-Balmaz, Michael Krauss, Giuseppe Marmo, Tomoki Ohsawa, Paul Skerritt, and Cornelia Vizman. This material is based upon work supported by the National Science Foundation under Grant No. DMS-1440140 while I was in residence at MSRI, during the Fall 2018 semester. In addition, I acknowledge financial support from the Leverhulme Trust Research Grant No. 2014-112, and from the London Mathematical Society Grant No. 31633 (Applied Geometric Mechanics).

\end{document}